\documentclass[aps,prd,reprint,superscriptaddress]{revtex4-1}

\usepackage{graphicx}
\usepackage{amsmath}
\usepackage{subfigure} 

\begin{document}

\title{Efficient Antihydrogen Detection in Antimatter Physics by Deep Learning}

\author{P. Sadowski}
\affiliation{Department of Computer Science, University of California, Irvine, CA 92617, USA}

\author{B. Radics}
\email[]{bradics@phys.ethz.ch}
\affiliation{Atomic Physics Laboratory, RIKEN, Saitama 351-0198, Japan}
\affiliation{ETH Zurich, Institute for Particle Physics, CH-8093 Z\"urich, Switzerland}

\author{Ananya}
\affiliation{Department of Computer Science, University of California, Irvine, CA 92617, USA}

\author{Y. Yamazaki}
\affiliation{Atomic Physics Laboratory, RIKEN, Saitama 351-0198, Japan}

\author{P. Baldi}
\affiliation{Department of Computer Science, University of California, Irvine, CA 92617, USA}

\begin{abstract}
Antihydrogen is at the forefront of antimatter research at the CERN Antiproton Decelerator. Experiments aiming to test the fundamental CPT symmetry and antigravity effects require the efficient detection of antihydrogen annihilation events, which is performed using highly granular tracking detectors installed around an antimatter trap. Improving the efficiency of the antihydrogen annihilation detection plays a central role in the final sensitivity of the experiments. We propose deep learning as a novel technique to analyze antihydrogen annihilation data, and compare its performance with a traditional track and vertex reconstruction method. We report that the deep learning approach yields significant improvement, tripling event coverage while simultaneously improving performance by over 5\% in terms of Area Under Curve (AUC).
\end{abstract}

\maketitle


\section{Introduction}
In recent years numerous experiments have successfully trapped \cite{Alpha_NatPhys2011, Gabrielse_PRL2012}, or formed a beam of, antihydrogen atoms \cite{Kuroda_NatComm2014}. The general aim of these experiments is to perform spectroscopic measurements in order to precisely compare the properties of the antihydrogen atom to that of the hydrogen atom, thus enabling a direct test of the CPT symmetry. Other experiments aim at testing the effects of gravity using antimatter \cite{Aegis_NatComm2014, GBar_Hyp, Alpha_antig}. Production of antihydrogen is performed either with antiproton and positron plasmas injected into cryogenic multi-ring electrode traps where three-body recombination takes place ($\bar{\mathrm{p}} + 2e^{+} \rightarrow \bar{\mathrm{H}} + e^{+}$), or by the charge exchange processes between positronium, antiproton, and antihydrogen ($\mathrm{Ps}^{(*)} + \bar{\mathrm{p}} \rightarrow \bar{\mathrm{H}} + e^{-}$ and $\bar{\mathrm{H}} + \mathrm{Ps} \rightarrow \bar{\mathrm{H}}^{+} + e^{-}$). In most of the cases, an antihydrogen event is detected by its annihilation signature whereby several charged pions are emitted from the annihilation vertex \cite{Hori03}. These are distinguished from cosmic or instrumental background events using the hit multiplicity and the infered positions of the tracks and annihilation vertex. 

Previous experiments have detected trapped antimatter annihilation events using tracking and vertex-finding approaches \cite{AthenaDet, AlphaDet, AtrapDet, AegisDet, BresciaDet1, BresciaDet2, AMT_RSI_15}, predominantly relying on algorithms adopted from High-Energy Physics (HEP). However, antimatter trap experiments are not designed for high-resolution and high-efficiency tracking performance; the necessary layers of vacuum chamber walls, multi-ring electrodes, and thermal isolation limits the resolution of the detector. Therefore, the vertex-reconstruction approach may not be optimal for detecting antihydrogen event signals.

In this work, we present a machine learning approach for analyzing data from antimatter trap experiments that does not require explicit vertex reconstruction. In particular, we propose the use of deep learning, essentially large neural networks, to identify  antihydrogen annihilation signatures in the low-level detector data. This approach has the advantage of being able to exploit subtle and previously-unrecognized characteristics in the data, as demonstrated in Higgs-search and Higgs-decay studies from HEP \cite{baldi2014searching,Higgs,baldidarkmatter15}. We train our system using Monte Carlo simulated data of antihydrogen-like signal and antiproton background annihilation events, and compare the performance to a traditional track and vertex reconstruction method. Although the results presented in this paper are obtained using a specific antimatter trap and detector environment, the approach is general and can be applied to other environments.

\section{\label{sec:Experiment}Monte Carlo simulation and annihilation event reconstruction}
Signal and background annihilation events were simulated with the ASACUSA experimental setup \cite{Kuroda_NatComm2014}. The primary antiproton annihilation process, the trap environment, and the Asacusa Micromegas Tracker (AMT) detector \cite{AMT_RSI_15} were simulated with the Geant4 toolkit \cite{Geant4}. The Monte Carlo data includes the full material budget of the cryogenic vacuum trap to account for the dominant effect of multiple Coulomb scattering of pions while emerging from the annihilation events and subsequently traversing the surrounding materials \cite{Shibata}. Hits produced by the simulated tracks from annihilation events are recorded and saved for analysis according to the AMT detector details. Briefly, the AMT detector consists of two, half-cylinder, curved micro-strip pattern gaseous detector layers using Micromegas technology  \cite{Giomataris96, cazaux2014detecteur}, and a single, full-cylinder layer of plastic trigger scintillator bars sandwiched in between the Micromegas layers, segmented into eight bars. The Micromegas single hit resolution was simulated with Garfield++, Magboltz, and Heed programs \cite{Heed, Garfieldpp} for the response of charged pions of $p \simeq 100$ MeV/c momentum and was found to be much better ($\sigma_{\mathrm{hit}} \simeq 250$ $\mu\mathrm{m}$) than the amount of smearing of the pion track positions ($\sigma_{\mathrm{Coul.}} \simeq 2$ mm) due to Coulomb scattering when the pions reach the AMT detector layers. Using the detected hits, the reconstructed vertex position resolution was estimated to be $\sigma_{vx} \simeq 1$ cm. The vertex position resolution may be used as a design parameter to discriminate antihydrogen signal events from antiproton or other background events. However, in numerous antimatter experiments aiming at trapping or forming a beam of antihydrogen the material budget is not optimized for annihilation vertex resolution but for antiproton and positron trapping performance.

Antihydrogen experiments trap charged particles (antiprotons and positrons) on the multi-ring electrode trap axis. The continuous annihilation of trapped antiprotons on the residual gas atoms produces background annihilation events. When a neutral antihydrogen atom is produced as a result of recombination of an antiproton with a positron, it may escape the electric trapping potentials and annihilate on the inner wall of the multi-ring electrodes. However, the difference in the annihilation signature of antiprotons or antihydrogen is only due to the annihilation vertex position. Therefore, in order to simulate antihydrogen signal events, we generated antiprotons to annihilate on the multi-ring electrode material, and to simulate antiproton background events annihilation process on residual gas atoms was used to generate events at the central axis of the trap. This scenario reflects realistic conditions of antihydrogen production observed in experiments \cite{Kuroda_NatComm2014}.

\begin{figure}
\hfill
\subfigure{\includegraphics[width=\linewidth]{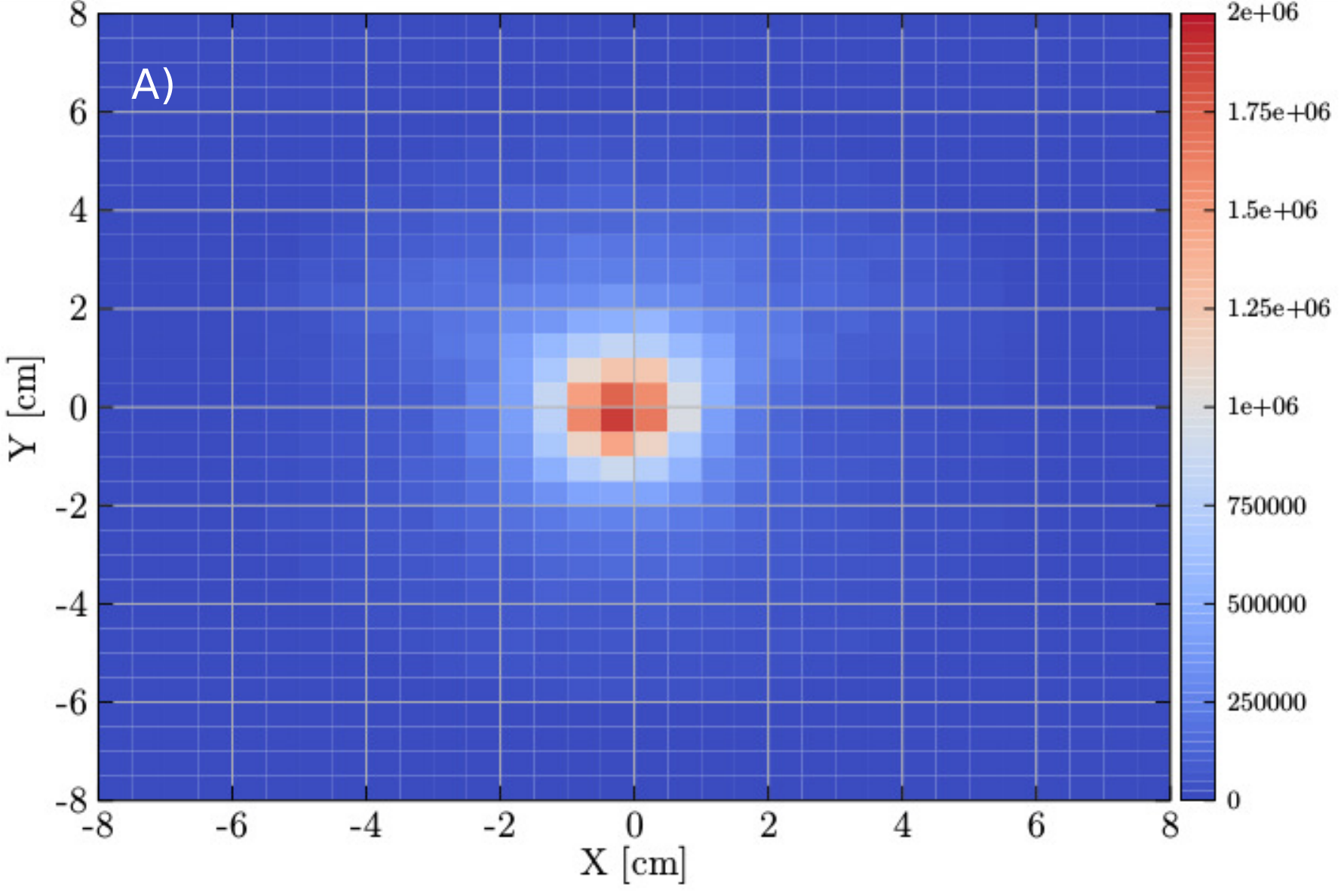}}
\hfill
\subfigure{\includegraphics[width=\linewidth]{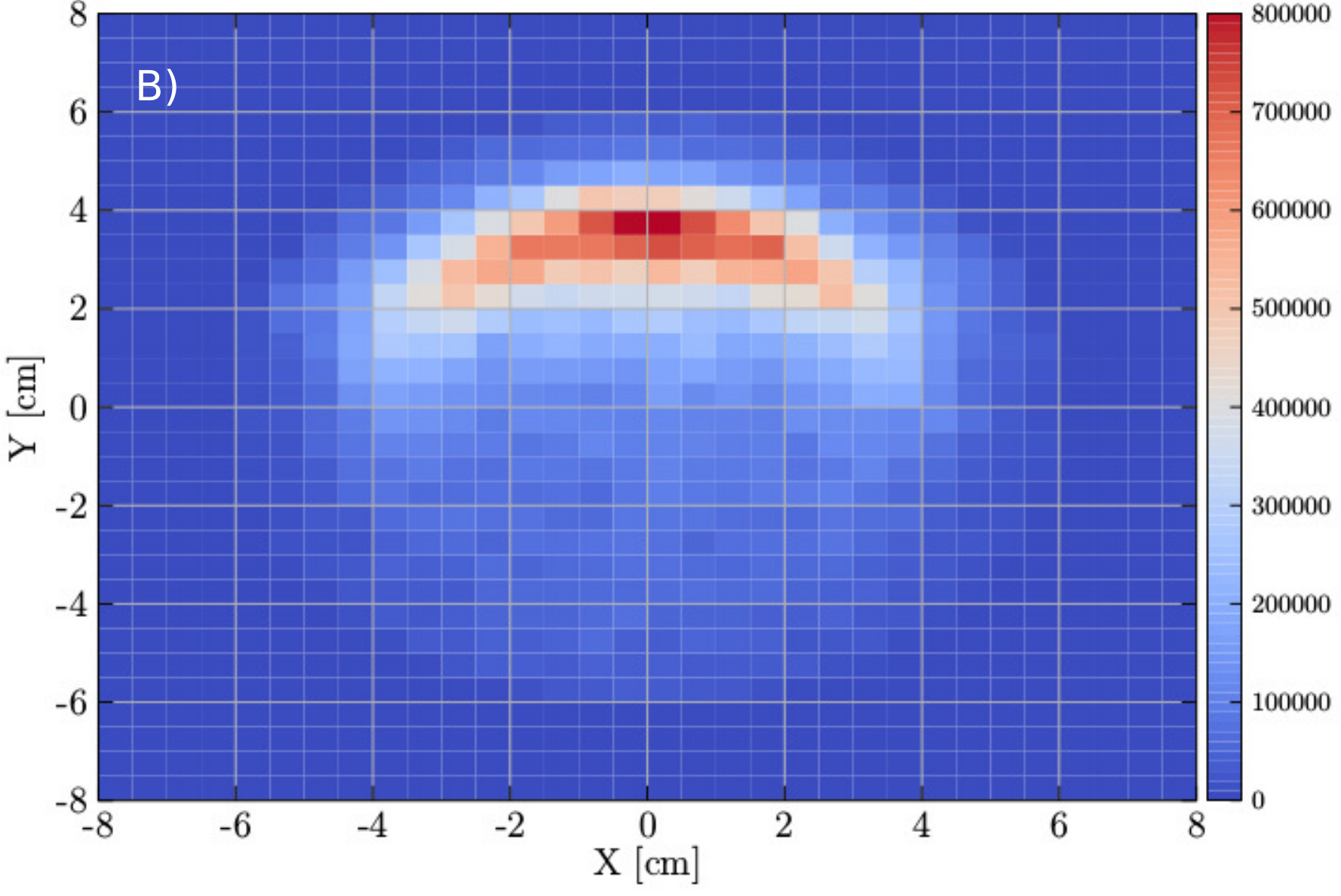}}
\hfill
\caption{Reconstructed annihilation vertex positions using the traditional track and vertex finding method for antiproton background events (A) and antihydrogen-like signal events on the multi-ring electrod wall at $R=4$ cm (B). The color scale indicates the number of vertices found at a given position.}
\label{fig:XYPos}
\end{figure}

In a typical track and vertex reconstruction algorithm, the recorded hit data is processed as follows: the single detector channels are searched for entries above threshold, and neighboring channels are iteratively clustered to form hits. The detected hits are then used to form and fit tracks, using a Kalman-filtering algorithm \cite{Kalmanfilter}. After fitting all hit pair combinations in an event, the track candidates are filtered by their compatibility with the known antiproton cloud position. Finally, the selected track candidates are paired and their three-dimensional point of closest approach position is assigned as the vertex position. The result of such a reconstruction approach are illustrated in Fig~\ref{fig:XYPos} for the case of antihydrogen-like signals (vertices distributed on the round multi-ring electrode walls) and antiproton background events (vertices confined to the center of the trap). In the ASACUSA experiment, tracking detector layers are only present in the upper half of the trap \cite{AMT_RSI_15}, which is reflected in the distribution of reconstructed vertices. The shape of the distribution also reflects the smearing of the position by 1 cm due to the Coulomb scattering of the charged pions in the trap material.

Event classification is based on the radial position of the reconstructed vertex, $R=\sqrt{X^2+Y^2}$, but the classes are not perfectly separable because low resolution results in heavy overlap in the distributions of $R$ for signal and background. Thus, a probabilistic model is used to classify the events. In this work, we use a simple neural network model with a single input $R$, one hidden layer of 100 hyperbolic tangent units, and a logistic output unit that returns a probability of an event being an antihydrogen annihilation. The cluster finding algorithm, track and vertex reconstruction algorithms, and neural network are all tuned from simulations and experimental test data.

\section{\label{sec:dl}Deep Learning Approach}
We propose a deep learning approach in which a deep neural network model is trained to classify annihilations directly from raw detector data. This automated, end-to-end learning strategy can potentially identify discriminative information in the raw data that is typically discarded in the vertex-reconstruction algorithm.

\begin{figure}
\centering
\includegraphics[width=\linewidth]{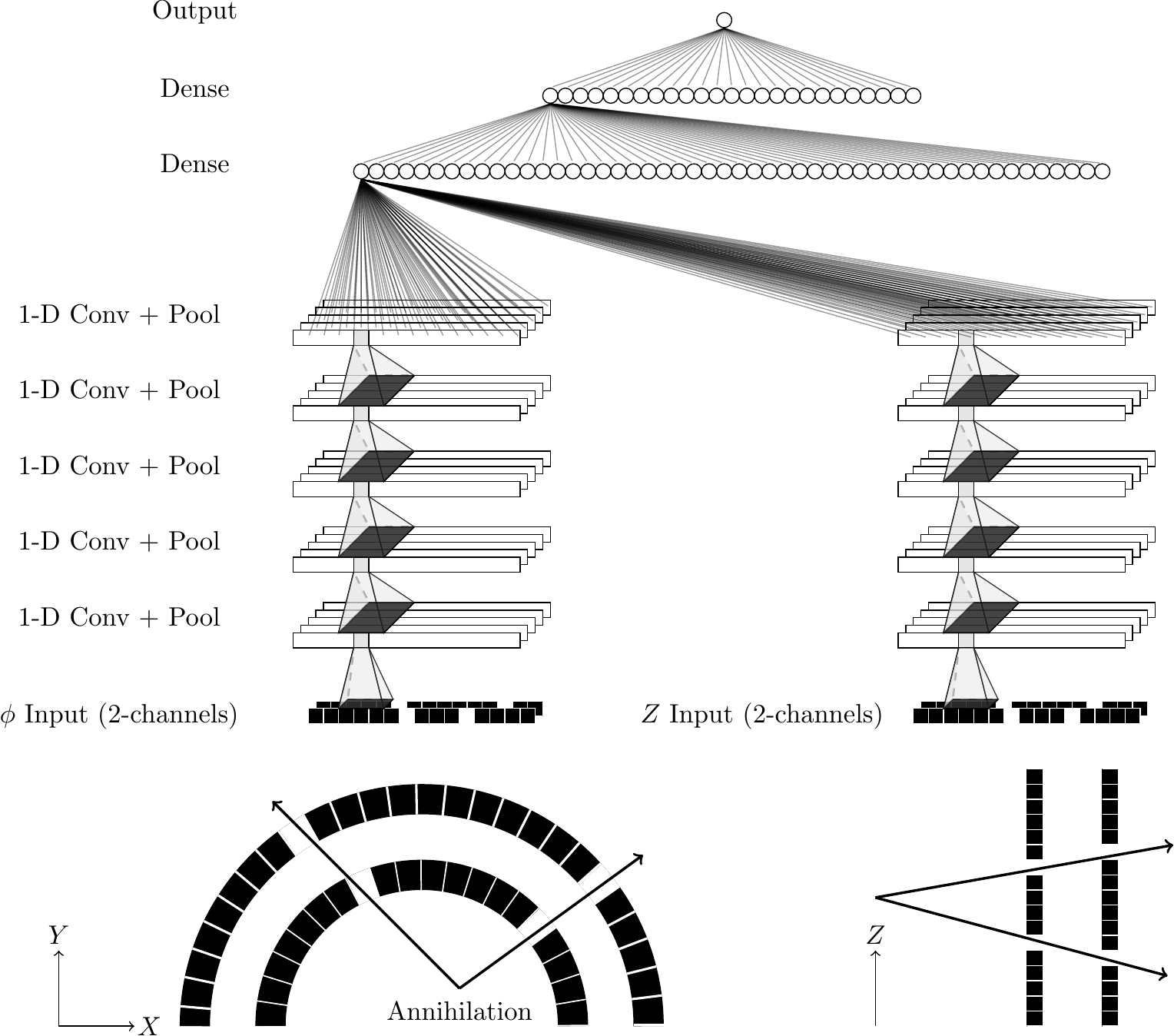}
\caption{Deep neural network architecture consisting of two separate pathways that process data from the micro-strips placed along the azimuthal and axial dimensions. In each pathway, the inner and outer detector layers are treated as two input channels and processed by a sequence of 1-D convolutional layers. Higher in the network, the information is merged using dense fully-connected layers.}
\label{fig:deeparch}
\end{figure}

Two design choices were made regarding the neural network architecture in order to address the geometry of the detector: (1) data from micro-strips situated along the azimuth ($\phi$) and axial ($Z$) dimensions are processed along separate pathways that are then merged higher in the architecture; and (2) 1-D convolution and max-pooling layers account for translational invariances in these dimensions. The raw data consists of 1,430 binary values from the two detector layers and two micro-strip orientations: 246 inner and 290 outer azimuthal strips, and 447 inner and 447 outer axial strips. The inner and outer detector layers are treated as two associated input 'channels' in the convolutional neural network, where the outer azimuthal layer is downsampled with linear interpolation to match the dimensionality of the inner layer. A diagram of this architecture is shown in Figure \ref{fig:deeparch}.

\section{\label{sec:results}Results}

One million annihilation events of each type were simulated and randomly divided into training (60\%), validation (20\%), and test subsets (20\%). Because the particles emitted by annihilations often escape undetected, the vertex finding (VF) algorithm failed on 75\% of these events: 20\% did not result in any detector hits; 7\% had a hit in only one detector; and 48\% had hits in both detectors, but a vertex could not be reconstructed. Vertex reconstruction failed because either (1) there were not enough distinct hits to infer the presence of at least two tracks, or (2) the point of closest approach of the reconstructed tracks was greater than the threshold value of 1 cm. Thus, a direct performance comparison was only possible on the 25\% of the test set events for which the VF algorithm succeeded, even though the deep learning approach provides a prediction for every event.

\begin{figure}
\centering
\vspace{1mm}
\includegraphics[width=\linewidth]{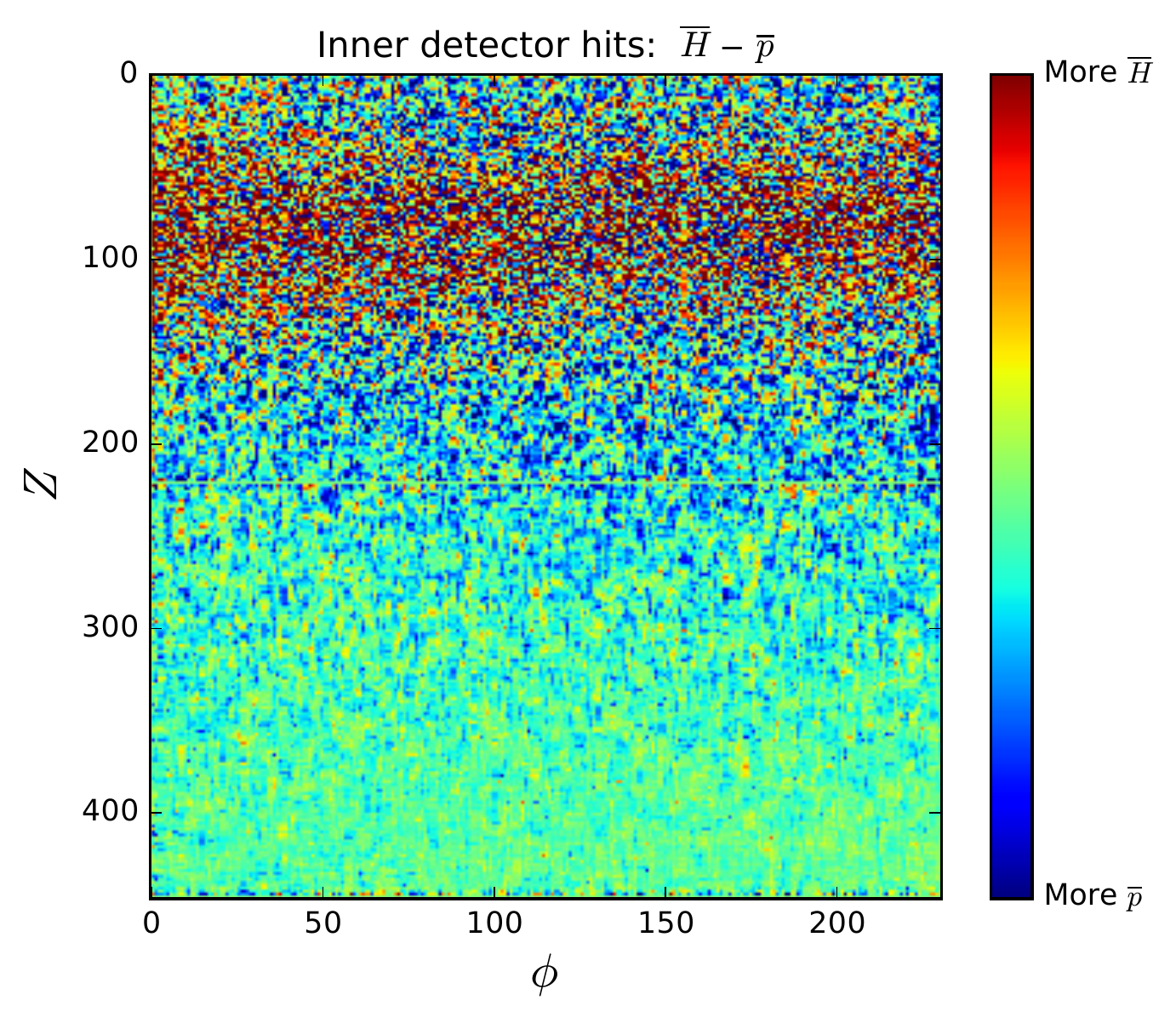}
\caption{Differential heatmap showing the relative density of hits from antihydrogen and antiproton annihilations in the Monte Carlo data, in which annihilations were simulated at $Z = 100$ cm. The true axial distribution of the annihilations is unknown, so in order to train the classifier to be translationally invariant, the data was augmented by translating each example by a random distance in $Z$.}
\label{fig:heatmap}
\end{figure}

Due to the complexity of the non-neutral plasma physics and atomic scattering processes, there is no physics model to our knowledge that would be able to predict the true axial distribution of the antihydrogen atoms in realistic experimental conditions. Therefore, the classifier should be invariant to translations in the axial dimension. This was achieved by augmenting the training data, in which simulated antihydrogen annihilations occured at $Z = 100$ cm (see Figure \ref{fig:heatmap}), by randomly translating each event during training by $t$ `pixels' in the $Z$ dimension, where $t$ is sampled from the uniform distribution $t \sim U(-100,347)$. The same augmentation was used on the validation and test sets to evaluate performance, but the performance was very similar on non-augmented data indicating that the network learned to focus on $Z$-invariant features.

Various neural network architectures were trained in order to optimize generalization performance on the validation set. The best architecture consists of five 1-D convolutional layers with kernel sizes 7-3-3-3-3 (the size of the receptive fields for neurons in each layer), channel sizes 8-16-32-64-128 (the number of distinct feature detectors in each layer), and rectified linear activation~\cite{icml2010_NairH10}. In order to account for translational invariance, each convolution layer is followed by a max-pooling layer with pool size 2 and stride length 2. The flattened representations from the two pathways are then concatenated and followed by two fully-connected layers of 50 and 25 rectified linear units, then a single logistic output unit with relative entropy error so that the output can be interpreted as a probability. During training, $50\%$ dropout was used in the top two dense layers to reduce overfitting~\cite{srivastava_dropout_2014,baldidropout14}. The model weights were initialized from a scaled normal distribution as suggested by He et al.~\cite{he_delving_2015}, then trained using the Adam optimizer~\cite{kingma2014adam} ($\beta_1=0.9, \beta_2=0.999, \epsilon=1e-08$) with mini-batch updates of size 100 and a learning rate that was initialized to $0.0001$ and decayed by 1\% at the end of each epoch. Training was stopped when the validation objective did not improve within a window of three epochs. The models were implemented in {\sc Keras}~\cite{chollet_keras_2015} and {\sc Theano}~\cite{theano2016}, and trained on a cluster of Nvidia Titan Black processors.

\begin{figure}
\centering
\includegraphics[width=\linewidth]{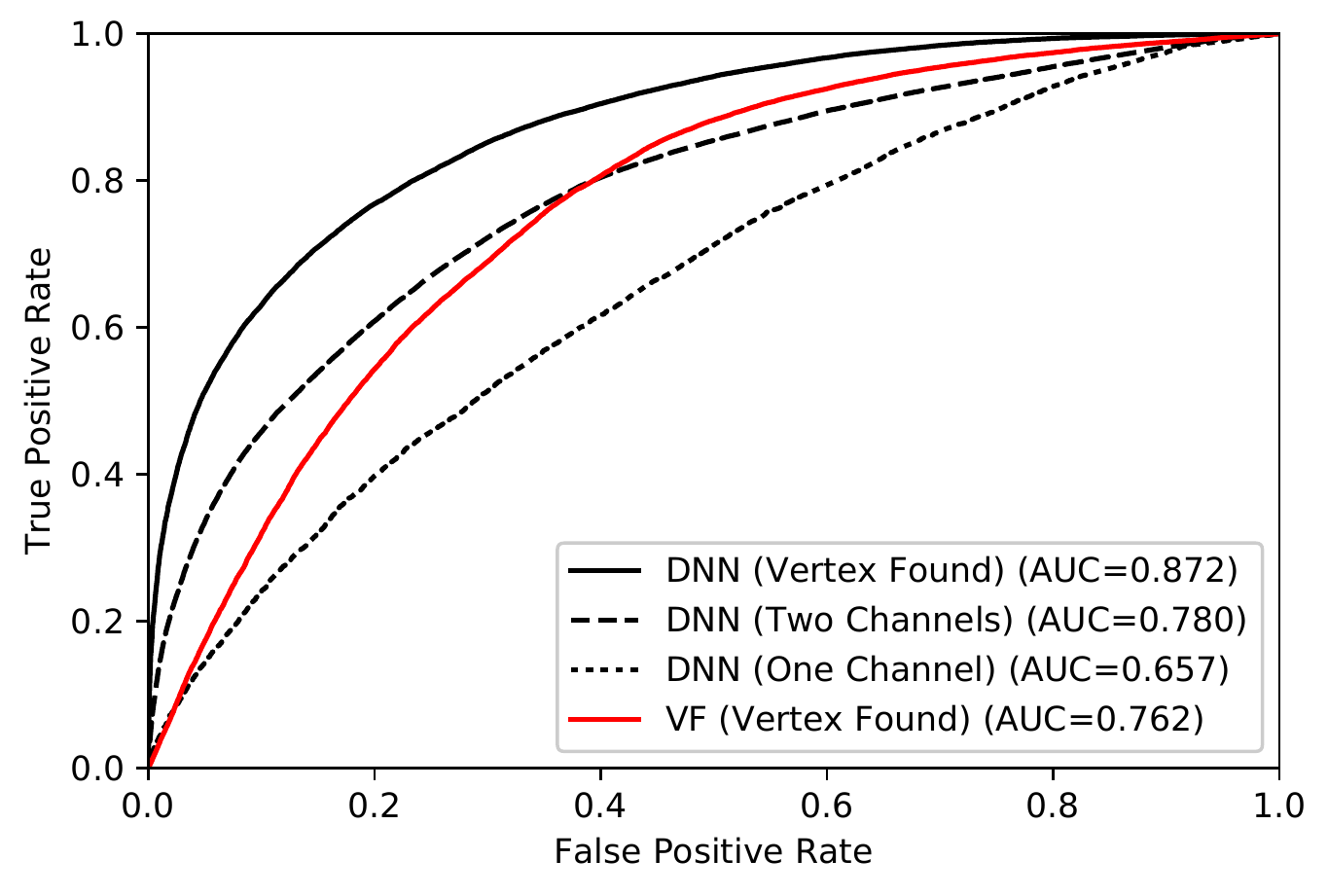}
\caption{ROC curves for the vertex finding algorithm (VF) and a deep convolutional neural network (DNN) on the subset of events for which an annihilation vertex can be reconstructed (`Vertex Found,' 25\% of all events). Also shown are the curves for DNN predictions on the subset of events in which both the inner and outer detector channels are hit, but no vertex could be reconstructed (48\% of all events), and on events in which only one channel is hit (7\% of all events). }
\label{fig:roc}
\end{figure}

Performance comparisons were made by plotting the Receiver Operating Characteristic (ROC) curve and calculating the Area Under the Curve (AUC). Figure \ref{fig:roc} shows that the deep neural network classifier outperforms the VF approach by a large margin on the test events for which a vertex could be reconstructed (87\% vs. 76\% AUC). Remarkably, the deep neural network achieves 78\% AUC on a disjoint set of events for which a vertex could \emph{not} be reconstructed, and 82\% AUC on the union set containing all events for which both detector layers were hit (not shown). This effectively triples the event coverage --- to 73\% of all events --- while \emph{simultaneously} improving the AUC by more than 5\%. These results clearly demonstrate that useful information contained in the raw data is being discarded by the VF algorithm, and that the deep learning approach is able to use this information to improve detection efficiency.

\begin{figure}
\centering
\includegraphics[width=\linewidth]{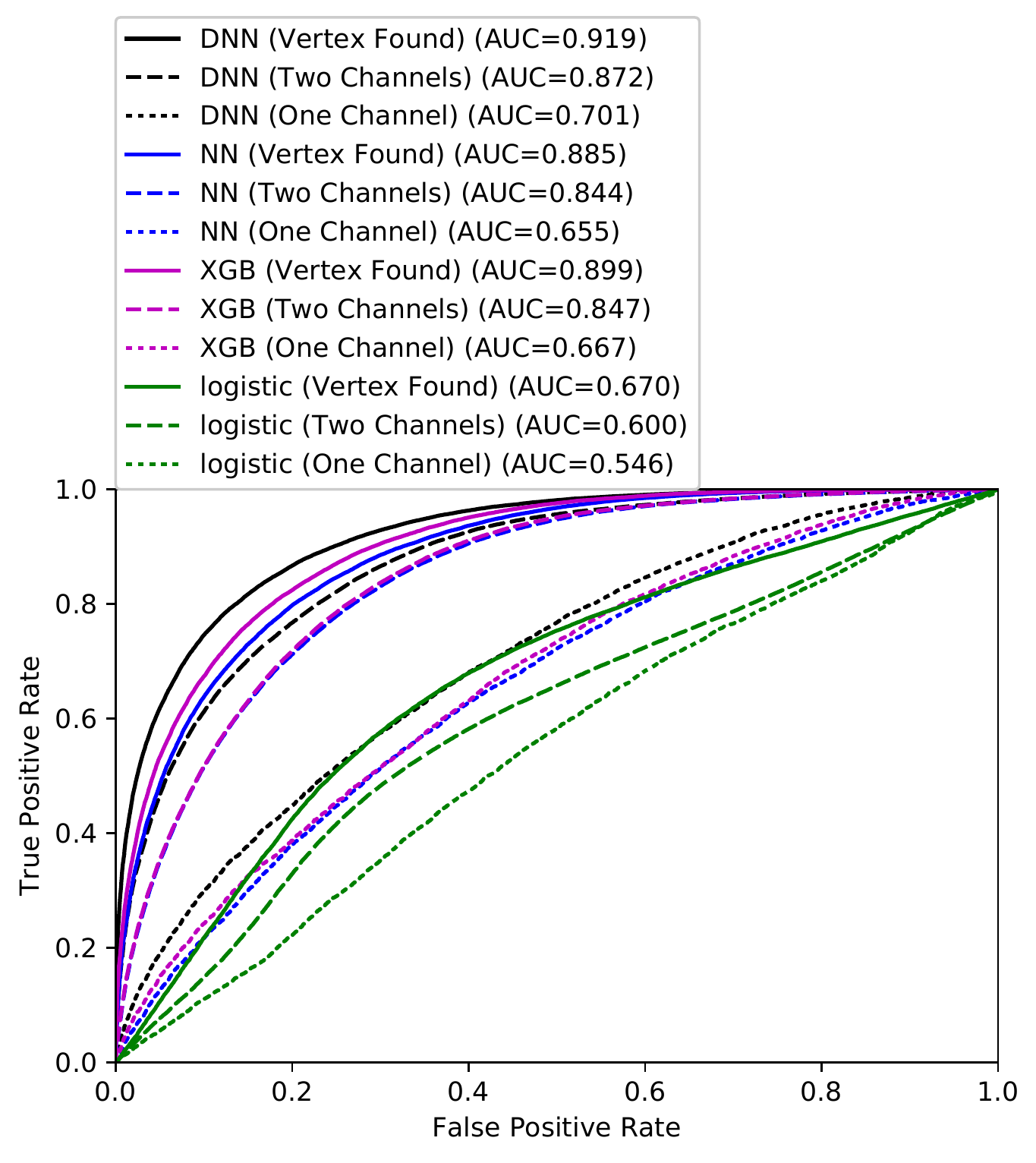}
\caption{ROC curves for four classification models on the non-augmented data set with annihilations fixed at $Z=100$: the deep convolutional architecture (DNN) from Figure \ref{fig:roc}, a shallow neural network with a single hidden layer of 2000 units (NN), XGBoost (XGB), and a logistic regression model. Predictions are made on the three event subsets: events for which a vertex can be reconstructed (Vertex Found), events in which both detector channels are hit but no vertex can be reconstructed (Two Channels), and events in which only one channel is hit (One Channel). The discriminative capability of the logistic regression model is predominately a result of the $Z$-dependence shown in Figure \ref{fig:heatmap}.
}
\label{fig:roc_depth}
\end{figure}

Additional experiments evaluated the advantage of deep learning compared to `shallow' machine learning with boosted decision trees and neural networks with only a single hidden layer. These experiments were performed on the non-augmented data, where the convolution and pooling layers provide \emph{less} of an advantage for the DNN. Figure \ref{fig:roc_depth} shows that XGBoost~\cite{chen2016} and shallow neural networks both perform better than the VF algorithm using the raw detector data (89\% and 90\% AUC on the same test events), but not as well as the DNN (92\% AUC). The shallow network had 2000 hidden rectified linear units and was trained with  70\% dropout in the hidden layer and the ADAM optimizer ($\alpha=0.0001, \beta_1=0.9, \beta_2=0.999, \epsilon=1e-8$); the hidden layer size, dropout probability, and learning rate were optimized based on validation performance. The XGBoost classifier had 100 trees with a maximum depth of 100 and minimum child weight of 10, and was trained with $\eta=0.1$, L2 regularization factor of 0.1, and feature sampling rate 0.5.

\section{\label{sec:Conclusion}Conclusion}
In summary, we report the first application of deep learning for the identification of antihydrogen annihilation events using realistic Monte Carlo simulations of both antihydrogen-like signals and antiproton background events. Moreover, we address the scenario in which classification must be translationally-invariant in the axial dimension to address systematic uncertainty in the experiment. The results demonstrate significant performance improvements compared to traditional approaches that are based on track and vertex reconstruction. Vertex reconstruction necessarily discards statistical information when fitting tracks, and the approach fails completely on all but 25\% of events. In the deep learning approach, the end-to-end neural network model extracts additional statistical information from the raw data, improving AUC from 76\% to 87\% on the same event subset. It also has far better coverage, achieving 82\% AUC on the 73\% of events in which both detector layers were hit. 

While this study was performed for the ASACUSA experiment, the deep learning approach can easily be deployed to other ongoing antimatter experiments with different instruments. Furthermore, the ability to detect complex signals in high-dimensional data without relying on explicit track reconstruction potentially offers a new direction in the design of future detectors.

Finally, the data used in this paper, comprising two million events, are publicly available from the machine learning in physics web portal: \url{http://mlphysics.ics.uci.edu}.
\\

\begin{acknowledgments}
The work of BR was supported by the Grant-in-Aid for Specially Promoted Research (no. 24000008) of the Japanese Ministry of Education, Culture, Sports, Science and Technology (Monbukakagu-sho), Special Research Projects for Basic Science of RIKEN, Pioneering Project of RIKEN. The work of PB and PS was in part supported by NSF grants 
IIS-1321053 and IIS-1550705, as well as an NVIDIA Corporation hardware award, to PB.
\\
\end{acknowledgments}

\bibliography{Bibliography}

\end{document}